\documentclass[aps,prd,twocolumn,showpacs]{revtex4-1}

\usepackage{amssymb}
\usepackage{graphicx}
\usepackage{amsmath}
\usepackage{amsbsy}
\usepackage{amsthm}
\usepackage{bbm}
\usepackage{bm}
\usepackage{epsfig}
\usepackage{epstopdf}
\usepackage{dsfont}
\usepackage{hyperref}
\usepackage{soul}
\usepackage{color}

\newcommand{\tr}{\textnormal{Tr}}

\newcommand{\be}{\begin{equation}}
\newcommand{\ee}{\end{equation}}
\newcommand{\beq}{\begin{eqnarray}}
\newcommand{\eeq}{\end{eqnarray}}

\DeclareMathOperator{\atanh}{atanh}

\DeclareMathOperator{\polylog}{Li}

\DeclareMathOperator{\Realpart}{Re}

\newcommand{\BbbR}{\mathbb{R}}
\newcommand{\BbbZ}{\mathbb{Z}}

\newcommand{\cavlength}{\delta}

\usepackage{geometry}
\begin{document}

\title{Kinematic entanglement degradation of fermionic cavity modes}
\date{January 2012}

\author{Nicolai Friis}
\email{pmxnf@nottingham.ac.uk}
\author{Antony R. Lee}
\email{pmxal3@nottingham.ac.uk}
\author{David Edward Bruschi}
\email{pmxdeb@nottingham.ac.uk}
\author{Jorma Louko}
\email{jorma.louko@nottingham.ac.uk}
\affiliation{School of Mathematical Sciences,
University of Nottingham,
University Park,
Nottingham NG7 2RD,
United Kingdom}
\begin{abstract}
We analyse the entanglement and the nonlocality of a
$(1+1)$-dimensional
massless Dirac field confined to a cavity on a worldtube that
consists of inertial and uniformly accelerated segments,
for small accelerations but arbitrarily long travel times.
The correlations between the accelerated field modes and the modes
in an inertial reference cavity are periodic in the durations of
the individual trajectory segments, and degradation of the
correlations can be entirely avoided by fine-tuning the
individual or relative durations of the segments.
Analytic results for selected trajectories are presented.
Differences from the corresponding bosonic correlations are identified
and extensions to massive fermions are discussed.
\end{abstract}

\maketitle

\section{\label{sec:intro}Introduction}

One of the fundamental problems in the emerging field of relativistic
quantum information is the degradation of correlations caused
by accelerated motion.  Studies of uniform acceleration in Minkowski
spacetime (see
\cite{alsingmilburn03,fuentesschullermann05,alsingfuentesschullermanntessier06,bruschiloukomartinmartinezdraganfuentes10,martinmartinezfuentes11,friiskoehlermartinmartinezbertlmann11}
for a small selection and \cite{MartinMartinez:2011mw} for a recent
review) have revealed significant differences in the degradation that
occurs for bosonic and fermionic fields.  There are in particular
clear qualitative differences in the bosonic versus fermionic
particle-antiparticle entanglement swapping
\cite{martinmartinezfuentes11} and in the infinite acceleration
residual entanglement and
nonlocality~\cite{friiskoehlermartinmartinezbertlmann11}.

The analyses of uniform acceleration mentioned above
involve two ingredients that make it difficult to compare the
theoretical predictions to experimentally realisable situations.
The first is that while the uniformly-accelerated
observers are considered to be pointlike and perfectly localised on a
trajectory of prescribed acceleration, the field excitations are
nevertheless usually treated as delocalised field modes of plane wave type,
normalised in the sense of Dirac rather than Kronecker deltas.
This may seem a technicality,
perhaps remediable by use of appropriate
wave packets~\cite{bruschiloukomartinmartinezdraganfuentes10},
but at present it appears unexplored how localised
observers would in practice perform measurements to probe
the correlations in the delocalised states.

The second concern lies in the time evolution of the correlations. An
inertial trajectory in Minkowski space is stationary, in the sense
that it is the integral curve of a Minkowski time translation Killing
vector. A~uniformly-accelerated trajectory is also stationary, in the
sense that it is the integral curve of a boost Killing vector. However, the
combined system of the two trajectories is not stationary, as
the two Killing vectors do not commute. For example, in the
$(1+1)$-dimensional setting there is a unique moment
at which the two trajectories are
parallel, and the trajectories may or may not intersect depending on
their relative spatial location.  Yet the analyses mentioned above
regard the correlations between observers on the two trajectories as
stationary and the relative location of the trajectories as
irrelevant, observing just that the spacetime has a quadrant causally
disconnected from the uniformly-accelerated worldline and noting that
the field modes confined in this quadrant are inaccessible to the
accelerated observer. While the acceleration horizon that is
responsible for this inaccessibility may be seen as the basis of the
Unruh effect~\cite{unruh,wald-smallbook}, the horizon exists
only if the uniform acceleration persists from the
asymptotic past to the asymptotic future. In this setting it is not
clear how to address motion on trajectories that remain uniformly
accelerated only up to the moment at which localised observers might
make their measurements on the quantum state.

Both of these concerns have been recently addressed by studying
correlations between field modes of a real scalar field confined in
two \emph{cavities\/}, one inertial and the other undergoing motion
that consists of segments of inertial motion and uniform
acceleration~\cite{bruschifuenteslouko11}. In this setting the field
modes are spatially localised in the cavities, and the acceleration
effects can be localised in time by taking the initial and final
segments of the accelerated cavity to be inertial. It was found that
the entanglement is affected by the acceleration, and in
$(1+1)$-dimensional spacetime the mass of the field has a strong
effect on the qualitative behaviour on the entanglement. For a
massless Dirichlet field the entanglement is periodic in the durations
of the individual trajectory segments, so that entanglement
degradation can be entirely avoided by fine-tuning the durations of
the individual segments; further, in the small acceleration limit the
degradation can also be avoided by fine-tuning the relative lengths of
the inertial and accelerated segments.

In this paper we shall undertake the first steps of investigating
fermionic entanglement in accelerated cavities by adapting the scalar
field analysis of \cite{bruschifuenteslouko11} to a Dirac
fermion. Conceptually, one new issue with fermions is that the
presence of positive and negative charges allows a broader range of
initial Bell-type states to be considered. Another conceptual issue is
that in a fermionic Fock space the
entanglement between the cavities can be characterised
not just by the negativity but also
by the violation of the Clauser-Horne-Shimony-Holt (CHSH) version of
Bell's inequality~\cite{chsh69,horodecki-rpm95}, physically interpretable as nonlocality. New technical issues arise from the
boundary conditions that are required to keep the fermionic field
confined in the cavities.

We focus this paper on a massless fermion in $(1+1)$ dimensions. In
this setting another new technical issue arises from a zero mode that
is present in the cavity under boundary conditions that may be
considered physically preferred. This zero mode needs to be
regularised in order to apply usual Fock space techniques.

We shall find that the entanglement behaviour of the massless Dirac
fermion is broadly similar to that found for the massless scalar
in~\cite{bruschifuenteslouko11}, in particular in the periodic
dependence of the entanglement on the durations of the individual
accelerated and inertial segments, and in the property that
entanglement degradation caused by accelerated segments can be
cancelled in the leading order in the small acceleration expansion by
fine-tuning the durations of the inertial segments. We shall however
find that the charge of the fermionic excitations has a quantitative
effect on the entanglement, and there is in particular interference
between excitations of opposite charge.

We begin in Sec.~\ref{sec:staticcavity} by quantising a massless
Dirac field in a static cavity and in a uniformly-accelerated cavity
in $(1+1)$ dimensional Minkowski spacetime. We pay special attention to
the boundary conditions that are required for maintaining unitarity
and to the regularisation of a zero mode that arises under an arguably
natural choice of the boundary conditions.
Section \ref{sec:bogo transformation} develops the Bogoliubov
transformation technique for grafting inertial and uniformly
accelerated trajectory segments, presenting the general building block
formalism and giving detailed results for a trajectory where initial
and final inertial segments are joined by one uniformly accelerated
segment. The evolution of initially maximally entangled states is
analysed in Sec.~~\ref{sec:fermionic state trafo}, and the results
for entanglement are presented in
Sec.~\ref{sec:degradation of entanglement}.
A one-way-trip travel scenario, in which the accelerated cavity undergoes both
acceleration and deceleration, is analysed in Sec.~\ref{sec:onewaytrip}. Section
\ref{sec:conclusion} presents a brief discussion and concluding
remarks.

We use units in which $\hbar=c=1$.
Complex conjugation is denoted by an asterisk
and Hermitian conjugation by a dagger.
$O(x)$ denotes a quantity for which
$O(x)/x$ is bounded as $x\to0$.

\section{\label{sec:staticcavity}Static cavity}

In this section we quantise the massless Dirac field in an inertial cavity
and in a uniformly accelerated cavity, establishing the notation
and conventions for use in the later sections.

\subsection{\label{subsec:inertialcavity}Inertial cavity}

Let $(t,z)$ be standard Minkowski coordinates in
\mbox{$(1+1)$} dimensional Minkowski space,
and let $\eta_{\mu\nu}$ denote the Minkowski metric,
$ds^2 = \eta_{\mu\nu} \, dx^\mu \, dx^\nu = -dt^2 + dz^2$.
The massless Dirac equation reads
\begin{align}
i\,\gamma^{\mu}\partial_{\mu}\,\psi=\,0\ ,
\label{eq:massless Dirac eq}
\end{align}
where the $4\times4$ matrices $\gamma^{\mu}$ form the usual Dirac-Clifford algebra,
$\left\{\gamma^{\mu},\gamma^{\nu}\right\}\,=\,2\,\eta^{\mu\nu}$.
A~standard basis of plane wave solutions reads
\begin{align}
\psi_{\omega,\epsilon,\sigma}(t,z)
\,=\,
A_{\omega,\epsilon,\sigma}
\,e^{-i\omega(t-\epsilon z)}\,U_{\epsilon,\sigma}
\, ,
\end{align}
where $\omega\in\BbbR$,
$\epsilon\in\{1,-1\}$,
$\sigma\in\{1,-1\}$,
the constant spinors
$U_{\epsilon,\sigma}$
satisfy
\begin{subequations}
\begin{align}
\alpha_{3} U_{\epsilon,\sigma}
&= \epsilon U_{\epsilon,\sigma} \ ,
\\
\gamma^{5} U_{\epsilon,\sigma}
&= \sigma U_{\epsilon,\sigma} \ ,
\\
U_{\epsilon,\sigma}^{\dagger}
U_{\epsilon^{\prime},\sigma^{\prime}}
& =
\delta_{\epsilon\epsilon^{\prime}}\delta_{\sigma\sigma^{\prime}}\, ,
\label{eq:spinor normalization condition}
\end{align}
\end{subequations}
and $A_{\omega,\epsilon,\sigma}$ is a normalisation constant.
Physically, $\omega$ is the
frequency with respect to the Minkowski time, the eigenvalue
$\epsilon$ of the operator $\alpha_{3}=\gamma^{0}\gamma^{3}$
indicates whether the solution is
a right-mover ($\epsilon=1$) or a left-mover
\mbox{($\epsilon=-1$)}, and $\sigma$ is the eigenvalue of the
helicity/chirality operator
$\gamma^{5}=i\gamma^{0}\gamma^{1}\gamma^{2}\gamma^{3}$~\cite{srednicki-book}.
The right-handed $(\sigma=+1)$ and left-handed $(\sigma=-1)$
solutions are decoupled because
(\ref{eq:massless Dirac eq}) does not contain a mass term.

We encase the field in the inertial cavity
$a \le z \le b$, where $a$ and $b$ are positive
parameters satisfying $a<b$.
The inner product reads
\begin{align}
\left(\,\psi_{(1)},\psi_{(2)}\right)
\,=\,
\int\limits_{a}^{b}\!dz\,\psi_{(1)}^{\dagger}\,\psi_{(2)}
\ ,
\label{eq:mink-ip}
\end{align}
where the integral is
evaluated on a surface of constant~$t$.
To ensure unitarity of the time evolution, so that
the inner product (\ref{eq:mink-ip}) is conserved in time, the Hamiltonian
must be defined as a self-adjoint operator by introducing suitable boundary conditions at
$z=a$ and $z=b$~\cite{reebk2,bonneauetal}.
We specialise to boundary conditions that do not
couple right-handed and left-handed spinors.
For concreteness, we consider from now on
only left-handed spinors,
and we drop the index~$\sigma$.
The analysis for right-handed spinors is similar.

We seek the eigenfunctions of the Hamiltonian in the form
\begin{align}
\psi_{\omega}(t,z)\,=\,A_\omega\,e^{-i\omega(t-z)}\,U_{+}
\,+\,B_\omega
\,e^{-i\omega(t+z)}\,U_{-}\, ,
\label{eq:in-psiraw}
\end{align}
where $A_\omega$ and $B_\omega$ are complex-valued constants.
It would be mathematically possible to maintain unitarity by
allowing probability to flow out
through one of the cavity walls and
instantaneously reappear at the other wall;
physically, this would mean that the spatial surface is considered
to be a circle, possibly with one marked point.
However, we wish to regard the spatial surface as a genuine interval
with two spatially separated endpoints, and we hence specialise
to boundary conditions that ensure
vanishing of the probability current independently at each wall.
The boundary condition on the eigenfunctions thus reads
\begin{align}
\left(\,\bar{\psi}_{\omega}\gamma^{3}\psi_{\omega'}\,\right)_{z=a}
\,=\,
0
\,=\,
\left(\,\bar{\psi}_{\omega}\gamma^{3}\psi_{\omega'}\,\right)_{z=b}
\ ,
\label{eq:in-bc}
\end{align}
where $\bar\psi=\psi^{\,\dagger}\gamma^{0}$.

Following the procedure of~\cite{reebk2,bonneauetal}, we find from
(\ref{eq:in-psiraw}) and (\ref{eq:in-bc}) that the self-adjoint
extensions of the Hamiltonian are specified by two independent phases,
characterising the phase shifts of reflection from the two walls.  We
encode these phases in the parameters $\theta \in [0,2\pi)$ and
$s\in[0,1)$, which can be understood respectively as the normalised
sum and difference of the two phases. The quantum theories then fall
into two qualitatively different cases, the generic case
$0<s<1$ and the special case $s=0$.

In the generic case $0<s<1$, the orthonormal eigenfunctions are
\begin{subequations}
\label{eq:in-psis}
\begin{align}
\psi_{n}(t,z)
& =
\frac{e^{-i\omega_{\!n}(t-z+a)}\,U_{+}\,+\,e^{i\theta}
e^{-i\omega_{\!n}(t+z-a)}\,U_{-}}{\sqrt{2\cavlength}}
\, ,
\label{eq:in-psis-psi}
\\
\omega_{n}
&= \frac{(n+s)\pi}{\cavlength}
\ ,
\end{align}
\end{subequations}
where $n\in\BbbZ$ and $\cavlength := b-a$.
Note that $\omega_{n} \ne0$ for all~$n$,
and positive (respectively negative)
frequencies are obtained for $n\ge0$ ($n<0$).
A~Fock space quantisation can be performed in
a standard manner~\cite{srednicki-book}.

The special case $s=0$ corresponds to assuming that the two walls are
of identical physical build. In this case $\omega_{n} \ne0$ for
$n\ne0$ but $\omega_{0}= 0$. It follows that a Fock quantisation can
proceed as usual for the $n\ne0$ modes, but $n=0$ is a zero mode
that does not admit a Fock space quantisation. In what follows we
consider the $s=0$ quantum theory to be defined by first quantising
with $s>0$ and at the end taking the limit $s\to0_+$.  All our
entanglement measures will be seen to remain well defined in this
limit.

\subsection{Uniformly accelerated cavity}

We consider a cavity whose ends move
on the worldlines
$z = \sqrt{a^2 + t^2}$
and
$z = \sqrt{b^2 + t^2}$, where the notation is as above
The proper accelerations of the ends are $1/a$ and~$1/b$ respectively,
and the cavity as a whole is static in the sense that
it is dragged along the boost Killing vector
$\xi := z\partial_t + t\partial_z$.
At $t=0$ the accelerated cavity overlaps precisely
with the inertial cavity of Sec.~\ref{subsec:inertialcavity}.

Coordinates convenient for the accelerated cavity are the
Rindler coordinates $(\eta,\chi)$,
defined in the quadrant $z > |t|$ by
\begin{align}
t=\chi\sinh\eta
\ , \
z=\chi\cosh\eta\ ,
\label{eq:mink-in-rindler}
\end{align}
where $0<\chi<\infty$ and $-\infty<\eta<\infty$.
The metric reads $ds^2 = -\chi^2 d\eta^2 + d\chi^2$.
The cavity is at $a \le \chi \le b$,
and the boost Killing vector along which
the cavity is dragged takes the form $\xi = \partial_\eta$.

In Rindler coordinates the massless Dirac equation (\ref{eq:massless Dirac eq})
takes the form
\cite{birrelldavies,mcmahonalsingembid06}
\begin{align}
i\,\partial_{\eta}\,{\psi}(\eta ,\chi)\,=\,\left(\,-i\,\alpha_{3}(\chi\,\partial_{\chi}
\,+\,\tfrac{1}{2})\,\right)\,{\psi}(\eta ,\chi)\ ,
\label{eq:dirac eq in rindler coords}
\end{align}
and the inner product for a field encased in
the accelerated cavity reads
\begin{align}
\bigl(\,{\psi}_{(1)},{\psi}_{(2)}\bigr)
\,=\,
\int\limits_{a}^{b} d\chi\,{\psi}_{(1)}^{\dagger}\,{\psi}_{(2)}
\ ,
\label{eq:rindler-ip}
\end{align}
where the integral is evaluated on a surface of constant~$\eta$.
Working as in
Sec.~\ref{subsec:inertialcavity},
we find that the orthonormal energy eigenfunctions are
\begin{subequations}
\label{eq:rind-psis}
\begin{align}
{\widehat{\psi}}_{n}(\eta ,\chi)
&=
\frac{\displaystyle e^{-i\Omega_{n}\eta}\left(\left(\frac{\chi}{a}\right)^{i\Omega_{n}}U_{+}+
e^{i\theta}\left(\frac{\chi}{a}\right)^{-i\Omega_{n}}U_{-}\right)}{\sqrt{2\chi\ln(b/a)}} ,
\\
\Omega_{n}
&=
\frac{(n\,+\,s)\pi}{\ln(b/a)}\ ,
\label{eq:rindler solutions frequency}
\end{align}
\end{subequations}
where $n\in\BbbZ$.
The parameters $\theta$ and $s$ have the same meaning and values as above:
we consider the microphysical build of the cavity
walls not to be affected by their acceleration.
For $s\ne0$ a Fock space quantisation can be performed in a standard manner.
For $s=0$ the mode $n=0$ is again a zero mode, and we consider the
$s=0$ quantum theory to be defined as the limit $s\to0_+$.

\section{\label{sec:bogo transformation}Grafting trajectory segments}

We now turn to a cavity whose
trajectory consists of inertial and uniformly accelerated segments.

\begin{figure}[t]
\begin{center}
\includegraphics[width=0.45\textwidth]{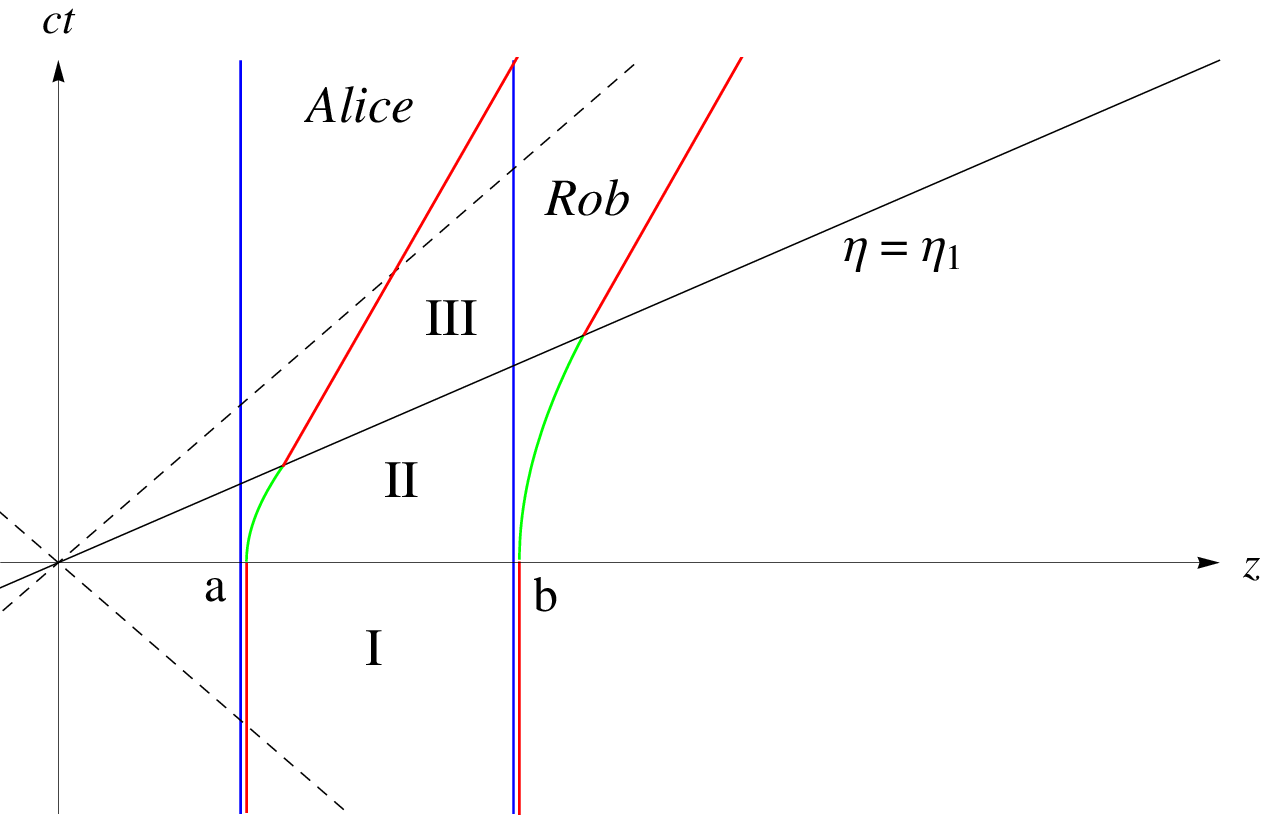}
\caption{
Space-time diagram of cavity motion is shown.
Rob's cavity is at rest initially (Region~$\mathrm{I}$),
then undergoes a period of uniform acceleration from $t=0$ to $\eta=\eta_{1}$
(Region $\mathrm{I\!I}$) and is thereafter
again inertial (Region $\mathrm{I\!I\!I}$).
Alice's cavity overlaps with
Rob's cavity in Region $\mathrm{I}$
and remains inertial throughout.}
\label{fig:bogobogoboxes}
\end{center}	
\end{figure}

The prototype cavity configuration is shown in Fig.~\ref{fig:bogobogoboxes}.
Two cavities, referred to as Alice and Rob, are initially
inertial and in the configuration described
in Sec.~\ref{subsec:inertialcavity}.
At $t=0$, Rob's cavity begins to accelerate to the right,
following the Killing vector $\xi = \partial_\eta$.
The acceleration ends at Rindler time $\eta = \eta_1$,
and the duration of the acceleration in proper time measured at the
centre of the cavity is
$\tau_1 := \tfrac12 (a+b)\eta_1$.
We refer to the three segments of Rob's trajectory as
Regions $\mathrm{I}$, $\mathrm{I\!I}$ and~$\mathrm{I\!I\!I}$.
Alice remains inertial throughout.

We shall discuss the evolution in Rob's cavity in two steps:
first from Region $\mathrm{I}$ to Region $\mathrm{I\!I}$
and then from Region $\mathrm{I\!I}$ to Region $\mathrm{I\!I\!I}$.
We then use the evolution to relate the operators and the vacuum of Region
$\mathrm{I}$ to those in Region~$\mathrm{I\!I\!I}$.

\subsection{Region $\mathrm{I}$ to Region $\mathrm{I\!I}$}
\label{subsec:I-to-II}

Consider the Dirac field in Rob's cavity. In Regions
$\mathrm{I}$ and $\mathrm{I\!I}$ we may expand the field using
the solutions (\ref{eq:in-psis}) and (\ref{eq:rind-psis}) respectively as
\begin{subequations}
\label{eq:regionsI-IIquantization}
\begin{align}
\mathrm{I}:\ \ \
\psi	&=	\sum\limits_{n\geq0}a_{n}\,\psi_{n}
\,+\,
\sum\limits_{n<0}b_{n}^{\dagger}\,\psi_{n}
\ ,
\label{eq:region I quantization}\\[2mm]
\mathrm{I\!I}:\ \ \
\psi	&=	\sum\limits_{m\geq0}\widehat{a}_{m}\,\widehat{\psi}_{m}
\,+\,
\sum\limits_{m<0}\widehat{b}_{m}^{\dagger}\,\widehat{\psi}_{m}
\ ,
\label{eq:region II quantization}
\end{align}
\end{subequations}
where the nonvanishing anticommutators are
\begin{subequations}
\label{eq:nonvanishing anticommutators}
\begin{align}
\mathrm{I}:\ \ \
&\left\{a_{m},a_{n}^{\dagger}\right\}=
\left\{b_{m},b_{n}^{\dagger}\right\}=\delta_{mn} \, ,
\\
\mathrm{I\!I}:\ \ \
&\left\{\widehat{a}_{m},\widehat{a}_{n}^{\dagger}\right\}=
\left\{\widehat{b}_{m},\widehat{b}_{n}^{\dagger}\right\}=\delta_{mn} \, .
\end{align}
\end{subequations}
Matching the expansions (\ref{eq:regionsI-IIquantization})
at $t=0$, we have
the Bogoliubov transformation
\begin{align}
\label{eq:change of basis}
\widehat{\psi}_{m}
= \sum\limits_{n}A_{mn}\,\psi_{n}
\ , \ \ \
\psi_{n}
= \sum\limits_{m}A_{mn}^{*}\,\widehat{\psi}_{m}\, ,
\end{align}
where the elements of the
Bogoliubov coefficient matrix $A = \left(A_{mn}\right)$
are given by
\begin{align}
A_{mn}\,=\,\bigl(\,\psi_{n},\widehat{\psi}_{m}\,\bigr)
\label{eq:bogo components from I to II}
\end{align}
and the inner product in (\ref{eq:bogo components from I to II})
is evaluated on the surface $t=0$.
Note that $A$ is unitary by construction.

We shall be working perturbatively in the limit
where the acceleration of Rob's cavity
is small. To implement this, we follow
\cite{bruschifuenteslouko11} and
introduce the dimensionless parameter
$h:=2\cavlength/(a+b)$, satisfying $0<h<2$.
Physically, $h$ is the product of the cavity's
length $\cavlength$ and the acceleration at
the centre of the cavity.
Expanding
(\ref{eq:bogo components from I to II})
in a Maclaurin series in~$h$,
we find
\begin{align}
A=A^{(0)}\,+\,A^{(1)}\,+\,A^{(2)}\,+\,{O}(h^{3})\ ,
\label{eq:bogo coefficient expansion}
\end{align}
where the superscript indicates
the power of $h$ and
the explicit expressions for
$A^{(0)}$, $A^{(1)}$ and $A^{(2)}$ read
\begin{subequations}
\label{eq:bogo coeff pert}
\begin{align}
A_{mn}^{(0)}
&=
\delta_{mn}\,,\label{eq:region I to II bogo order zero}
\\[1ex]
A_{nn}^{(1)}
&= 0 \,,
\\[1ex]
A_{mn}^{(1)}
&=
\frac{\bigl[(-1)^{m+n}-1\bigr](m+n+2s)}{2\pi^{2}(m-n)^{3}}\,h \,,
\ \ \ (m\ne n)
\\
A_{nn}^{(2)}
&= -
\left(\frac{1}{96}+\frac{\pi^{2}(n+s)^2}{240}\right) h^{2}\,,
\\[1ex]
A_{mn}^{(2)}
&=
\frac{\bigl[(-1)^{m+n}+1\bigr]}{8\pi^{2}(m-n)^{4}}
\bigl[(m+s)^{2}+3(n+s)^{2}
\notag
\\[.5ex]
&
\hspace{9ex}
+8(m+s)(n+s) \bigr]h^{2} \,.
\ \ \ (m\ne n)
\end{align}
\end{subequations}
The expressions (\ref{eq:bogo coeff pert}) show that the small $h$
expansion of $A_{mn}$ is not uniform in the indices, but we have
verified that the expansion maintains the unitarity of $A$
perturbatively to order $h^2$ and the products of the order $h$
matrices in the unitarity identities are unconditionally
convergent.

The perturbative unitarity of $A$ persists in the
limit $s\to0_+$. Had we set $s=0$ at the outset and dropped the zero
mode from the system by hand, the resulting truncated $A$ would not be
perturbatively unitary to order~$h^2$.

\subsection{Region $\mathrm{I}$ to Region $\mathrm{I\!I\!I}$}

In Region $\mathrm{I\!I\!I}$, we expand the Dirac field in Rob's cavity as
\begin{align}
\mathrm{I\!I\!I}:\ \ \
\psi	&=	\sum\limits_{n\geq0}\tilde{a}_{n}\,\tilde{\psi}_{n}
\,+\,
\sum\limits_{n<0}\tilde{b}_{n}^{\dagger}\,\tilde{\psi}_{n}
\ ,
\label{eq:region III quantization}
\end{align}
where the mode functions $\tilde{\psi}_{n}$ are as in
(\ref{eq:in-psis}) but $(t,z)$ are replaced by the
Minkowski coordinates $(\tilde{t},\tilde{z})$
adapted to the cavity's new rest frame,
with the surface $\tilde{t}=0$ coinciding with $\eta=\eta_1$.
The nonvanishing anticommutators are
\begin{align}
\mathrm{I\!I\!I}:\ \ \
&\left\{\tilde{a}_{m},\tilde{a}_{n}^{\dagger}\right\}=
\left\{\tilde{b}_{m},\tilde{b}_{n}^{\dagger}\right\}=\delta_{mn} \, .
\label{eq:III-anticommutators}
\end{align}
The Bogoliubov transformation between the
Region
$\mathrm{I}$
and Region
$\mathrm{I\!I\!I}$
modes can then be written as
\begin{align}
\label{eq:IIIchange of basis}
\tilde{\psi}_{m}
= \sum\limits_{n}\mathcal{A}_{mn}\,\psi_{n}
\ , \ \ \
\psi_{n}
= \sum\limits_{m}\mathcal{A}_{mn}^{*}\,\tilde{\psi}_{m}\, ,
\end{align}
where the coefficient matrix $\mathcal{A} = \left(\mathcal{A}_{mn}\right)$
has the decomposition
\begin{align}
\mathcal{A}
=
A^\dagger\,G(\eta_{1})\,A
\label{eq:bogo from I to III}
\end{align}
and $G(\eta_1)$
is the diagonal matrix whose diagonal elements are
\begin{align}
G_{nn}(\eta_1) =
\exp(i\,\Omega_{n}\,\eta_{1}) =: G_n
\,.
\label{eq:free Rindler evolution diagonal matrix}
\end{align}
The role of $G(\eta_{1})$ in (\ref{eq:bogo from I to III})
is to compensate for the phases that the modes $\widehat{\psi}_{m}$
develop between $\eta=0$ and $\eta=\eta_1$,
and the matrix $A^\dagger = A^{-1}$
in (\ref{eq:bogo from I to III})
arises from matching
Region
$\mathrm{I\!I}$
to Region
$\mathrm{I\!I\!I}$ at $\eta=\eta_1$.
Note that $\mathcal{A}$ is unitary by construction.

Expanding $\mathcal{A}$ in a Maclaurin series in $h$ as
\begin{align}
\mathcal{A}=\mathcal{A}^{(0)}\,+\,\mathcal{A}^{(1)}
\,+\,
\mathcal{A}^{(2)}\,+\,{O}(h^{3})\ ,
\label{eq:IIIbogo coefficient expansion}
\end{align}
where the superscript again indicates
the power of $h$, we obtain from
(\ref{eq:bogo coefficient expansion}) and (\ref{eq:bogo from I to III})
\begin{subequations}
\label{eq:region I to III bogos three orders}
\begin{align}
\mathcal{A}^{(0)}
&=
G(\eta_{1})
\,,
\label{eq:region I to III bogo order zero}
\\[1.5mm]
\mathcal{A}^{(1)}
&=
G(\eta_{1})\,A^{(1)} + \bigl(A^{(1)}\bigr)^\dagger G(\eta_{1})
\,,
\label{eq:region I to III bogo order one}
\\[1.5mm]
\mathcal{A}^{(2)}
&=
G(\eta_{1})\,A^{(2)}
+ \bigl(A^{(2)}\bigr)^\dagger G(\eta_{1})
+ \bigl(A^{(1)}\bigr)^\dagger G(\eta_{1}) \,A^{(1)}
\,.
\label{eq:region I to III bogo order two}
\end{align}
\end{subequations}
Note that
the diagonal elements of $\mathcal{A}^{(1)}$ are vanishing.
Unitarity of $\mathcal{A}$ implies the perturbative relations
\begin{subequations}
\begin{align}
0 &= G(\eta_{1})^* \mathcal{A}^{(1)}
+
\bigl(\mathcal{A}^{(1)}\bigr)^\dagger G(\eta_{1})
\,,
\label{eq:first order unitarity condition}
\\[1.5mm]
0 &= G(\eta_{1})^* \mathcal{A}^{(2)}
+
\bigl(\mathcal{A}^{(2)}\bigr)^\dagger G(\eta_{1})
+ \bigl(\mathcal{A}^{(1)}\bigr)^\dagger \mathcal{A}^{(1)}
\,,
\label{eq:second order unitarity condition}
\end{align}
\end{subequations}
which will be useful below.

\subsection{Operators and vacua}

We denote the Fock vacua in Regions
$\mathrm{I}$ and $\mathrm{I\!I\!I}$
by $\left|\,0\,\right\rangle$ and $\left|\,\tilde{0}\,\right\rangle$ respectively.
To relate the two,
we mimic the bosonic case \cite{fabbrinavarrosalas}
and make the ansatz
\begin{align}
\left|\,0\,\right\rangle\,=\,N e^W \left|\,\tilde{0}\,\right\rangle \, ,
\label{eq:vacuum trafo}
\end{align}
where
\begin{align}
W\,=\,\sum\limits_{p\ge 0,q<0}\,V_{pq}\,\tilde{a}_{p}^{\dagger}\,\tilde{b}_{q}^{\dagger}
\label{eq:W vacuum trafo}
\end{align}
and the coefficient matrix $V = \left(V_{pq}\right)$
and the normalisation constant $N$ are to be determined.
Note that the two indices of $V$ take values in disjoint sets.

It follows from
(\ref{eq:region I quantization}),
(\ref{eq:region III quantization})
and
(\ref{eq:IIIchange of basis})
that the creation and annihilation operators
in Regions
$\mathrm{I}$ and $\mathrm{I\!I\!I}$ are related by
\begin{subequations}
\begin{align}
n\geq0:\
a_{n} &=\left(\,\psi_{n},\psi\,\right)
= \sum\limits_{m\geq0}\tilde{a}_{m}\,\mathcal{A}_{mn}\,+\,\sum\limits_{m<0}\tilde{b}_{m}^{\dagger}\,\mathcal{A}_{mn}\ ,
\label{eq:bogo decomposition of a sub n full}
\\
n<0:\ b_{n}^{\dagger} &= \left(\,\psi_{n},\psi\,\right)
=
\sum\limits_{m\geq0}\tilde{a}_{m}\,\mathcal{A}_{mn}\,+\,\sum\limits_{m<0}\tilde{b}_{m}^{\dagger}\,\mathcal{A}_{mn}\ .
\label{eq:bogo decomposition of bdag sub n full}
\end{align}
\end{subequations}
Using (\ref{eq:vacuum trafo}) and~(\ref{eq:bogo decomposition of a sub n full}),
the condition $a_{n}\left|\,0\,\right\rangle=0$ reads
\begin{align}
\Biggl(\,
\sum\limits_{m\geq0}\tilde{a}_{m}\,\mathcal{A}_{mn}
+\sum\limits_{m<0}\tilde{b}_{m}^{\dagger}\,\mathcal{A}_{mn}
\Biggr)
e^{W} \left|\,\tilde{0}\,\right\rangle=0\,.
\label{eq:using vacuum definition}
\end{align}
From the anticommutators
(\ref{eq:III-anticommutators})
it follows that
\begin{subequations}
\label{eq:aW-commutators}
\begin{align}
& \left[\,W\,,\,\tilde{a}_{m}\,\right]\,=\,-\sum\limits_{q<0}V_{mq}\,\tilde{b}_{q}^{\dagger}\,,
\\
& \left[\,W\,,\,\left[\,W\,,\,\tilde{a}_{m}\,\right]\,\right]\,=\,0\,.
\end{align}
\end{subequations}
Using (\ref{eq:aW-commutators}) and the
Hadamard lemma,
\begin{align}
e^{A}\,Be^{-A}\,=\,B\,+\,\left[\,A\,,\,B\,\right]\,+\,\tfrac{1}{2}\left[\,A\,,\,\left[\,A\,,\,B\,\right]\,\right]\,+\,\ldots
,
\label{eq:hadamard lemma}
\end{align}
(\ref{eq:using vacuum definition}) reduces to
\begin{align}
\sum\limits_{m\geq0}\mathcal{A}_{mn}\,V_{mq}\,=\,-\,\mathcal{A}_{qn}
\ \ \ (n\,\geq\,0\,,\ q\,<\,0)\ .
\label{eq:first relation between V and A}
\end{align}
A similar computation shows that
the condition $b_{n}\,\left|\,0\,\right\rangle=0$
reduces to
\begin{align}
\sum\limits_{m<0}\mathcal{A}_{mn}^{*}\,V_{pm}\,=\,\,\mathcal{A}_{pn}^{*}
\ \ \ (n\,<\,0\,,\ p\,\geq\,0)\ .
\label{eq:second relation between V and A}
\end{align}

If the block of $\mathcal{A}$ where both indices are non-negative is
invertible, Eq.~(\ref{eq:first relation between V and A})
determines $V$ uniquely. Similarly, if the block of $\mathcal{A}$
where both indices are negative is invertible, 
Eq.~(\ref{eq:second relation between V and A}) determines $V$ uniquely. If
both blocks are invertible, it can be verified using unitarity of
$\mathcal{A}$ that the ensuing two expressions for $V$ are
equivalent. Working perturbatively in~$h$, the invertibility
assumptions hold, and using (\ref{eq:IIIbogo coefficient expansion})
and (\ref{eq:region I to III bogos three orders}) we find
\begin{align}
V\,=\,V^{(1)}\,+\,{O}(h^{2})
\label{eq:expansion of V}
\end{align}
where
\begin{align}
V_{pq}^{(1)}\,=\,-\mathcal{A}_{qp}^{(1)}\,G_{p}^{\,*}
\,=\,
\mathcal{A}_{pq}^{(1)*}\,G_{q}\,\ \ (p\ge 0, \ q<0).
\label{eq:leading order of V}
\end{align}
We shall show in
Section \ref{sec:fermionic state trafo} that the normalisation
constant $N$ has the small $h$ expansion
\begin{align}
N\,=\,1\,-\,\tfrac{1}{2}\sum\limits_{p,q}|V_{pq}|^{2}
\ +{O}(h^{3})\,.
\label{eq:normalization constant to order h squared}
\end{align}

\section{\label{sec:fermionic state trafo}Evolution of entangled states}

In this section we apply the results of Section \ref{sec:bogo transformation}
to the evolution of Bell-type quantum states between the two cavities which are initially maximally
entangled.
We shall work perturbatively to quadratic order in~$h$.\\

Focusing first on Rob's cavity only,
we write out in Sec.~\ref{subsec:vacuum and single particle states}
the Region $\mathrm{I}$ vacuum and the Region $\mathrm{I}$ states with a
single (anti-)particle in terms of Region $\mathrm{I\!I\!I}$
excitations on the Region $\mathrm{I\!I\!I}$ vacuum.
In Sec.~\ref{subsec:two mode entangled states} we address
an entangled state where one field mode is controlled by Alice and one by Rob.
In Sec.~\ref{subsec:particle antiparticle entangled states} we address
a state of the type analysed in \cite{martinmartinezfuentes11}
where the entanglement between Alice and Rob
is in the charge of the field modes.

\subsection{\label{subsec:vacuum and single particle states}Rob's cavity:
vacuum and single-particle states}

Consider the Region $\mathrm{I}$ vacuum $\left|\,0\,\right\rangle$ in
Rob's cavity.
We shall use (\ref{eq:vacuum trafo}) to express this state in terms of
Region $\mathrm{I\!I\!I}$ excitations over the
Region $\mathrm{I\!I\!I}$ vacuum~$\left|\,\tilde{0}\,\right\rangle$.\\

We expand the exponential in (\ref{eq:vacuum trafo}) as
\begin{align}	 		
e^W
&=
\mathds{1}\,+\,\sum\limits_{p,q}V_{pq}\,\tilde{a}_{p}^{\dagger}\,\tilde{b}_{q}^{\dagger}
\notag
\\[1.5mm]
&\hspace{5ex}
+ \tfrac{1}{2}\,\sum\limits_{p,q,i,j}V_{pq}\,V_{ij}\,
\tilde{a}_{p}^{\dagger}\,\tilde{b}_{q}^{\dagger}\,\tilde{a}_{i}^{\dagger}\,\tilde{b}_{j}^{\dagger}
\,+\,{O}(h^{3})\,.
\label{eq:expansion of expW}
\end{align}
We denote the Region $\mathrm{I\!I\!I}$ single-particle states by
\begin{align}
\left|\tilde{1}_{k}\right\rangle^{\!+}
:= \tilde{a}_{k}^{\dagger}\left|\,\tilde{0}\,\right\rangle
\label{eq:single particle state}
\end{align}
for $k\geq0$ and by
\begin{align}
\left|\tilde{1}_{k}\right\rangle^{\!-}
:= \tilde{b}_{k}^{\dagger}\left|\,\tilde{0}\,\right\rangle
\label{eq:single antiparticle state}
\end{align}
for $k<0$, so that the superscript $\pm$ indicates the sign of the charge. From (\ref{eq:expansion of expW}) we obtain
\begin{align}
e^{W}\,\left|\,\tilde{0}\,\right\rangle
&=
\left|\,\tilde{0}\,\right\rangle
\,+\,\sum\limits_{p,q}V_{pq}\,
\left|\,\tilde{1}_{p}\right\rangle^{+}\,\left| \tilde{1}_{q}\,\right\rangle^{-}			
\notag
\\[.5ex]
&\hspace{1ex}
-\tfrac{1}{2}\sum\limits_{p,q,i,j}V_{pq}V_{ij}(1-\delta_{pi})(1-\delta_{qj})
\times
\notag
\\[.5ex]
&\hspace{6ex}
\times
\left|\tilde{1}_{p}\right\rangle^{\!+}
\left|\tilde{1}_{i}\right\rangle^{\!+}
\left|\tilde{1}_{q}\right\rangle^{\!-}
\left|\tilde{1}_{j}\right\rangle^{\!-}
+{O}(h^{3})\,,
\label{eq:expansion of expW acting on tilde0}
\end{align}
where the ordering of the single-particle kets
encodes the ordering of the fermion creation operators.
It follows that the normalisation constant $N$ is given
by~(\ref{eq:normalization constant to order h squared}),
and (\ref{eq:vacuum trafo}) gives

\begin{align}
\left|\,0\,\right\rangle
&=
\Bigl(1\,-\,\tfrac{1}{2}\sum\limits_{p,q}|V_{pq}|^{2}\Bigr)
\left|\,\tilde{0}\,\right\rangle
\,+\,\sum\limits_{p,q}V_{pq}\,
\left|\,\tilde{1}_{p}\right\rangle^{+}\,\left| \tilde{1}_{q}\,\right\rangle^{-}			
\notag
\\[.5ex]
&\hspace{1ex}
-\tfrac{1}{2}\sum\limits_{p,q,i,j}V_{pq}V_{ij}(1-\delta_{pi})(1-\delta_{qj})
\times
\notag
\\[.5ex]
&\hspace{6ex}
\times
\left|\tilde{1}_{p}\right\rangle^{\!+}
\left|\tilde{1}_{i}\right\rangle^{\!+}
\left|\tilde{1}_{q}\right\rangle^{\!-}
\left|\tilde{1}_{j}\right\rangle^{\!-}
+{O}(h^{3})\,.
\label{eq:region I vacuum state to order h squared}
\end{align}\\

Consider then in Rob's cavity the state with exactly one
Region $\mathrm{I}$ particle,
$\left|1_{k}\right\rangle^{\!-} := {b}_{k}^{\dagger} \left|\,0\,\right\rangle$ for $k<0$
or
\mbox{$\left|1_{k}\right\rangle^{\!+} := {a}_{k}^{\dagger} \left|\,0\,\right\rangle$} for $k\ge0$.
Acting on the Region $\mathrm{I}$ vacuum (\ref{eq:region I vacuum state to order h squared})
by (\ref{eq:bogo decomposition of bdag sub n full})
and the Hermitian conjugate of~(\ref{eq:bogo decomposition of a sub n full}) respectively,
we find\\

\begin{widetext}
\begin{subequations}	
\label{eq:region I single particles to order h squared}
\begin{align}	
&
k<0:
\ \ \
\left|1_{k}\right\rangle^{\!-}
=
\sum\limits_{p,q}V_{pq}\mathcal{A}_{pk}\left|\tilde{1}_{q}\right\rangle^{\!-}+
\sum\limits_{m<0}\mathcal{A}_{mk}
\Biggl[
\Bigl(1-\tfrac{1}{2}\sum\limits_{p,q}|V_{pq}|^{2}\Bigr)
\left|\tilde{1}_{m}\right\rangle^{\!-}+
\sum\limits_{p,q}V_{pq}(1-\delta_{mq})
\left|\tilde{1}_{p}\right\rangle^{\!+}\!
\left|\tilde{1}_{q}\right\rangle^{\!-}\!
\left|\tilde{1}_{m}\right\rangle^{\!-}
\notag
\\[1mm]	
&\hspace{20ex}
- \tfrac{1}{2}\sum\limits_{p,q,i,j}\hspace*{-1mm}V_{pq}V_{ij}
(1\!-\!\delta_{pi})(1\!-\!\delta_{qj})(1\!-\!\delta_{mq})(1\!-\!\delta_{mj})
\left|\tilde{1}_{p}\right\rangle^{\!+}\!
\left|\tilde{1}_{i}\right\rangle^{\!+}\!
\left|\tilde{1}_{q}\right\rangle^{\!-}\!
\left|\tilde{1}_{j}\right\rangle^{\!-}\!
\left|\tilde{1}_{m}\right\rangle^{\!-}
\Biggr]
+{O}(h^{3})\,,
\label{eq:region I single antiparticle state to order h squared}
\\[4.0mm]
&
k>0: \ \ \
\left|1_{k}\right\rangle^{\!+}
=
-\sum\limits_{p,q}V_{pq}\mathcal{A}_{qk}^{*}
\left|\tilde{1}_{p}\right\rangle^{\!+}+
\sum\limits_{m\geq0}\mathcal{A}_{mk}^{*}
\Biggl[
\Bigl(1-\tfrac{1}{2}\sum\limits_{p,q}|V_{pq}|^{2}\Bigr)
\left|\tilde{1}_{m}\right\rangle^{\!+}
+\sum\limits_{p,q}V_{pq}\,(1-\delta_{mp})
\left|\tilde{1}_{m}\right\rangle^{\!+}\!
\left|\tilde{1}_{p}\right\rangle^{\!+}\!
\left|\tilde{1}_{q}\right\rangle^{\!-}
\notag\\[1mm]
&\hspace{20ex}
- \tfrac{1}{2}\sum\limits_{p,q,i,j}V_{pq}V_{ij}
(1\!-\!\delta_{pi})(1\!-\!\delta_{qj})(1\!-\!\delta_{mp})(1\!-\!\delta_{mi})
\left|\tilde{1}_{m}\right\rangle^{\!+}\!
\left|\tilde{1}_{p}\right\rangle^{\!+}\!
\left|\tilde{1}_{i}\right\rangle^{\!+}\!
\left|\tilde{1}_{q}\right\rangle^{\!-}\!
\left|\tilde{1}_{j}\right\rangle^{\!-}
\Biggr]
+{O}(h^{3})\,.
\label{eq:region I single particle state to order h squared}
\end{align}
\end{subequations}
\end{widetext}

\subsection{\label{subsec:two mode entangled states}Entangled two-mode states}

We wish to consider a Region $\mathrm{I}$ state where
one field mode is controlled by Alice and one by Rob.
Concretely, we take
\begin{subequations}
\label{eq:initial pure state rob both}
\begin{align}
\left|\,\phi^{\pm}_{\text{init}}\,\right\rangle_{AR+}
&=
\tfrac{1}{\sqrt{2}} \left(\,
\left|\,0_{\hat{k}}\,\right\rangle_{A}\left|\,0_{k}\,\right\rangle_{R}\,\pm\,
\left|\,1_{\hat{k}}\,\right\rangle_{A}^{\kappa}\left|\,1_{k}\,\right\rangle_{R}^{+}\,\right),
\label{eq:initial pure state rob particle}
\\
\left|\,\phi^{\pm}_{\text{init}}\,\right\rangle_{AR-}
&=
\tfrac{1}{\sqrt{2}} \left(\,
\left|\,0_{\hat{k}}\,\right\rangle_{A}\left|\,0_{k}\,\right\rangle_{R}\,\pm\,
\left|\,1_{\hat{k}}\,\right\rangle_{A}^{\kappa}\left|\,1_{k}\,\right\rangle_{R}^{-}\,\right),
\label{eq:initial pure state rob antiparticle}
\end{align}
\end{subequations}
where the subscripts $A$ and $R$ refer to the cavity and the
superscripts $\pm$ indicate whether the mode has positive or negative
frequency, so that
$\kappa =+$ for $\hat{k}\ge0$ and $\kappa =-$ for $\hat{k}<0$.
Furthermore, we consider the two particle basis state of the two mode Hilbert space,
corresponding to one excitation each in the modes $\hat{k}$ in Alice's cavity and~$k$ in
Rob's cavity, to be ordered as in~(\ref{eq:initial pure state rob both}). As pointed out in
Ref.~\cite{monteromartinmartinez11}, making such a choice can lead to ambiguities in the
entanglement. In our case, the ambiguity amounts to a relative phase shift of $\pi$, i.e.,
a sign change, in (\ref{eq:initial pure state rob both}), which does not affect the amount
of entanglement. In other words, the states (\ref{eq:initial pure state rob both})
are pure, bipartite, maximally entangled states of mode
$\hat{k}$ in Alice's cavity and mode $k$ in Rob's cavity.

We form the density matrix for each of the states~(\ref{eq:initial pure state rob both}),
express the density matrix in terms of Rob's Region $\mathrm{I\!I\!I}$ basis to order $h^2$ using
(\ref{eq:region I vacuum state to order h squared})
and~(\ref{eq:region I single particles to order h squared}),
and take the partial trace over all of Rob's modes except the reference mode~$k$.
All of Rob's modes except $k$ are thus regarded as environment,
to which information is lost due to the acceleration.
The relevant partial traces of Rob's matrix elements depend on the sign of the mode label~$k$.
For $k\geq\,0$, corresponding to~(\ref{eq:initial pure state rob particle}), we find
\begin{subequations}
\label{eq:par-tr k pos all}
\begin{align}
\tr_{\lnot k}\left|\,0_{k}\right\rangle\left\langle\,0_{k}\right|
&=
(1-f_{k}^{-})\left|\,\tilde{0}_{k}\right\rangle\left\langle\tilde{0}_{k}\right|
+f_{k}^{-}\left|\tilde{1}_{k}\right\rangle^{\!\!++\!\!}\!\left\langle\tilde{1}_{k}\right|,
\label{eq:par-tr k pos 0 0}
\\
\tr_{\lnot k}\left|\,0_{k}\right\rangle^{+}\!\!\!\left\langle1_{k}\right|
&=
\Bigl(G_{k}+\mathcal{A}_{kk}^{(2)}\Bigr)
\left|\,\tilde{0}_{k}\right\rangle^{+}\!\!\!\left\langle\tilde{1}_{k}\right|,
\label{eq:par-tr k pos 0 1+}
\\
\tr_{\lnot k}\left|1_{k}\right\rangle^{\!\!++\!\!}\!\left\langle1_{k}\right|
&=
(1-f_{k}^{+})
\left|\tilde{1}_{k}\right\rangle^{\!\!++\!\!}\!\left\langle\tilde{1}_{k}\right|+
f_{k}^{+}\left|\,\tilde{0}_{k}\right\rangle\left\langle\tilde{0}_{k}\right|,
\label{eq:par-tr k pos 1+ 1+}
\end{align}
\end{subequations}
where we have used (\ref{eq:first order unitarity condition})
and (\ref{eq:leading order of V})
and introduced the abbreviations
\begin{align}
f_{k}^{+}
:=
\sum\limits_{p\geq0}\bigl|\mathcal{A}_{pk}^{(1)}\bigr|^{2}
\,, \ \ \ \
f_{k}^{-}
:=
\sum\limits_{q<0}\bigl|\mathcal{A}_{qk}^{(1)}\bigr|^{2}
\,.
\end{align}
For $k<0$, corresponding to~(\ref{eq:initial pure state rob antiparticle}),
we find similarly
\begin{subequations}
\label{eq:par-tr k neg all}
\begin{align}
\tr_{\lnot k}\left|\,0_{k}\right\rangle\left\langle\,0_{k}\right|
&=
(1-f_{k}^{+})\left|\,\tilde{0}_{k}\right\rangle\left\langle\tilde{0}_{k}\right|
+f_{k}^{+}\left|\tilde{1}_{k}\right\rangle^{\!\!--\!\!}\!\left\langle\tilde{1}_{k}\right|,
\label{eq:par-tr k neg 0 0}
\\
\tr_{\lnot k}\left|\,0_{k}\right\rangle^{-}\!\!\!\left\langle1_{k}\right|
&=
\Bigl(G_{k}^{\,*}+\mathcal{A}_{kk}^{(2)*}\bigr)
\left|\,\tilde{0}_{k}\right\rangle^{-}\!\!\!\left\langle\tilde{1}_{k}\right|,
\label{eq:par-tr k neg 0 1-}
\\
\tr_{\lnot k}\left|1_{k}\right\rangle^{\!\!--\!\!}\!\left\langle1_{k}\right|
&=
(1-f_{k}^{-})
\left|\tilde{1}_{k}\right\rangle^{\!\!--\!\!}\!\left\langle\tilde{1}_{k}\right|+
f_{k}^{-}\left|\,\tilde{0}_{k}\right\rangle\left\langle\tilde{0}_{k}\right|.
\label{eq:par-tr k neg 1- 1-}
\end{align}
\end{subequations}


\subsection{\label{subsec:particle antiparticle entangled states}States
with entanglement between opposite charges}

We finally consider the Region $\mathrm{I}$ state
\begin{align}
\left|\,\chi^{\pm}_{\text{init}}\,\right\rangle_{AR}\,=\,
\tfrac{1}{\sqrt{2}}
\left(\,
\left|\,1_{k}\,\right\rangle_{A}^{\!+}\left|\,1_{k^{\prime}}\,\right\rangle_{R}^{\!-}\,\pm\,
\left|\,1_{k^{\prime}}\,\right\rangle_{A}^{\!-}\left|\,1_{k}\,\right\rangle_{R}^{\!+}\,\right),
\label{eq:initial particle antiparticle entangled state}
\end{align}
where the meaning of the subscripts and superscripts is as described for Eq.~(\ref{eq:initial pure state rob both}), indicating that
$k\geq0$ and $k^{\prime}<0$. In this state Alice and Rob each have access to both of the modes
$k$ and $k^{\prime}$, and
the entanglement is in the \emph{charge\/} of the field modes,
similarly to the states considered in~\cite{martinmartinezfuentes11}.

We form the reduced density matrix to order $h^2$
as in Sec.~\ref{subsec:two mode entangled states},
but now the partial tracing over Rob's modes excludes both mode $k$ and mode~$k^{\prime}$.
The relevant matrix elements take the form
\begin{subequations}
\label{eq:par-tr k,kprime all}
\begin{align}
&\tr_{\lnot k,k^{\prime}}\left|1_{k^{\prime}}\right\rangle^{\!\!--\!\!}\!\left\langle1_{k^{\prime}}\right|\ =\
f_{k^{\prime}}^{-}
\left|\,\tilde{0}_{k}\right\rangle^{\!\!+}\!\left|\,\tilde{0}_{k^{\prime}}\right\rangle^{\!\!--\!\!}\!
\left\langle\tilde{0}_{k^{\prime}}\right|^{+\!\!}\!\left\langle\tilde{0}_{k}\right|
\notag
\\[1.5mm]
&\ +\,\bigl(1\!-\!f_{k^{\prime}}^{-}\!-\!f_{k}^{-}\!+\!\bigl|\mathcal{A}_{kk^{\prime}}^{(1)}\bigr|^{2}\bigr)
\left|\,\tilde{0}_{k}\right\rangle^{\!\!+}\!\left|\,\tilde{1}_{k^{\prime}}\right\rangle^{\!\!--\!\!}\!
\left\langle\tilde{1}_{k^{\prime}}\right|^{+\!\!}\!\left\langle\tilde{0}_{k}\right|
\notag
\\[1.5mm]
&\ +\,\bigl(f_{k}^{-}\!-\!\bigl|\mathcal{A}_{kk^{\prime}}^{(1)}\bigr|^{2}\bigr)
\left|\,\tilde{1}_{k}\right\rangle^{\!\!+}\!\left|\,\tilde{1}_{k^{\prime}}\right\rangle^{\!\!--\!\!}\!
\left\langle\tilde{1}_{k^{\prime}}\right|^{+\!\!}\!\left\langle\tilde{1}_{k}\right|
\notag
\\[1.5mm]
&\ +\,\Bigl(\sum\limits_{q<0}G_{k}G_{k^{\prime}}^{\,*}\mathcal{A}_{qk}^{(1)*}\mathcal{A}_{qk^{\prime}}^{(1)}
\left|\,\tilde{0}_{k}\right\rangle^{\!\!+}\!\left|\,\tilde{0}_{k^{\prime}}\right\rangle^{\!\!--\!\!}\!
\left\langle\tilde{1}_{k^{\prime}}\right|^{+\!\!}\!\left\langle\tilde{1}_{k}\right|
\notag
\\
&\  \hspace{5ex}\,+\,h.c.\Bigr)\,,
\label{eq:par-tr k,kprime 1- 1-}
\\[1.5mm]
&\tr_{\lnot k,k^{\prime}}\left|1_{k}\,\right\rangle^{\!\!++\!\!}\!\left\langle1_{k}\,\right|\ =\
f_{k}^{+}
\left|\,\tilde{0}_{k}\right\rangle^{\!\!+}\!\left|\,\tilde{0}_{k^{\prime}}\right\rangle^{\!\!--\!\!}\!
\left\langle\tilde{0}_{k^{\prime}}\right|^{+\!\!}\!\left\langle\tilde{0}_{k}\right|
\notag
\\[1.5mm]
&\ +\,\bigl(1\!-\!f_{k^{\prime}}^{+}\!-\!f_{k}^{+}\!+\!\bigl|\mathcal{A}_{kk^{\prime}}^{(1)}\bigr|^{2}\bigr)
\left|\,\tilde{1}_{k}\right\rangle^{\!\!+}\!\left|\,\tilde{0}_{k^{\prime}}\right\rangle^{\!\!--\!\!}\!
\left\langle\tilde{0}_{k^{\prime}}\right|^{+\!\!}\!\left\langle\tilde{1}_{k}\right|
\notag
\\[1.5mm]
&\ +\,\bigl(f_{k^{\prime}}^{+}\!-\!\bigl|\mathcal{A}_{kk^{\prime}}^{(1)}\bigr|^{2}\bigr)
\left|\,\tilde{1}_{k}\right\rangle^{\!\!+}\!\left|\,\tilde{1}_{k^{\prime}}\right\rangle^{\!\!--\!\!}\!
\left\langle\tilde{1}_{k^{\prime}}\right|^{+\!\!}\!\left\langle\tilde{1}_{k}\right|
\notag
\\[1.5mm]
&\ -\,\Bigl(\sum\limits_{p\geq0}G_{k}G_{k^{\prime}}^{\,*}\mathcal{A}_{pk}^{(1)*}\mathcal{A}_{pk^{\prime}}^{(1)}
\left|\,\tilde{0}_{k}\right\rangle^{\!\!+}\!\left|\,\tilde{0}_{k^{\prime}}\right\rangle^{\!\!--\!\!}\!
\left\langle\tilde{1}_{k^{\prime}}\right|^{+\!\!}\!\left\langle\tilde{1}_{k}\right|
\notag
\\
&\  \hspace{5ex}\,+\,h.c.\Bigr)\,,
\label{eq:par-tr k,kprime  1+ 1+}
\\[1.5mm]
&\tr_{\lnot k,k^{\prime}}\left|1_{k}\,\right\rangle^{\!\!+-\!\!}\!\left\langle1_{k^{\prime}}\,\right|
=
\Bigl(
G_{k}^{\,*}G_{k^{\prime}}^{\,*}
\bigl|\mathcal{A}_{kk^{\prime}}^{(1)}\bigr|^{2}
\notag\\[1.5mm]
& \ \ +
\mathcal{A}_{kk}^{*}\mathcal{A}_{k^{\prime}k^{\prime}}^{*}
\Bigr)
\left|\,\tilde{1}_{k}\right\rangle^{\!\!+}\!\left|\,\tilde{0}_{k^{\prime}}\right\rangle^{\!\!--\!\!}\!
\left\langle\tilde{1}_{k^{\prime}}\right|^{+\!\!}\!\left\langle\tilde{0}_{k}\right|\,,
\label{eq:par-tr k,kprime  1+ 1-}
\end{align}
\end{subequations}
where in (\ref{eq:par-tr k,kprime  1+ 1-})
$\mathcal{A}_{kk}^{*}\mathcal{A}_{k^{\prime}k^{\prime}}^{*}$
is kept only to order $h^2$ in the small $h$ expansion
\begin{align}
\mathcal{A}_{kk}^{*}\mathcal{A}_{k^{\prime}k^{\prime}}^{*}\,=\,
G_{k}^{\,*}G_{k^{\prime}}^{\,*}+
G_{k^{\prime}}^{\,*}\mathcal{A}_{kk}^{(2)*}+
G_{k}^{\,*}\mathcal{A}_{k^{\prime}k^{\prime}}^{(2)*}+{O}(h^{3})\,.
\label{eq:Akk times Akprimekrprime expansion}
\end{align}

\section{\label{sec:degradation of entanglement}Entanglement degradation and nonlocality}

We are now in a position to study the entanglement and the nonlocality
of our states in Region~$\mathrm{I\!I\!I}$.

\subsection{\label{subsec:two mode entanglement}Entanglement of two-mode states}

Consider the states
$\left|\,\phi^{\pm}_{\text{init}}\,\right\rangle_{AR+}$ and
$\left|\,\phi^{\pm}_{\text{init}}\,\right\rangle_{AR-}
$~(\ref{eq:initial pure state rob both}),
in which
Alice and Rob control one mode each.
We shall quantify the entanglement by the negativity
\cite{vidalwerner02,audenaert-etal,plenio-virmani:review}
and the nonlocality by a possible violation of the
CHSH inequality~\cite{chsh69,horodecki-rpm95}.

The negativity $\mathcal{N}[\rho]$ is an entanglement monotone that
quantifies how strongly the partial transpose of a density operator
$\rho$ fails to be positive.  It is defined as the sum of the absolute
values of the negative eigenvalues $\lambda$ of $(\mathds{1}\otimes
T_{R})\rho$,
\begin{align}
\mathcal{N}[\rho]\,=\,\sum\limits_{\lambda<0}\,|\lambda|\ ,
\label{eq:negativity}
\end{align}
where $(\mathds{1}\otimes T_{R})$ denotes the transpose in one of the
two subsystems (which we have taken to be Rob without loss of
generality).  The negativity is a useful measure for our system
because all the entangled states that it fails to detect are
necessarily bound entangled, that is, these states cannot be
distilled~\cite{horo3-prl}, and a system with two fermionic modes
cannot be bound entangled.

We work perturbatively in~$h$.  The unperturbed part of
$(\mathds{1}\otimes T_{R})\rho_{AR\pm}^{\pm}$ has the triply
degenerate eigenvalue $\tfrac{1}{2}$ and the nondegenerate
eigenvalue~$-\tfrac{1}{2}$. In a perturbative treatment the positive
eigenvalues remain positive and the only correction to the negativity
comes from the perturbative correction to the negative eigenvalue.
A~straightforward computation using (\ref{eq:par-tr k pos all}) and
(\ref{eq:par-tr k neg all}) shows that the leading correction to the
negativity comes in order~$h^2$, and to this order the
negativity formula reads
\begin{align}
\mathcal{N}[\rho^{\pm}_{AR\pm}]\,=\,\tfrac{1}{2} \left( 1 - f_k \right)
\label{eq:negativity of rhoARplusminus}
\end{align}
where $f_{k} :=
f_{k}^{+}+f_{k}^{-}$.
$f_{k}$~can be expressed as
\begin{align}
f_{k}
&
=
\sum\limits_{p=-\infty}^{\infty}
\bigl|E_1^{k-p} - 1\bigr|^2
\bigl| A_{kp}^{(1)}\bigr|^{2}
\notag
\\[1ex]
& =
2\bigl[
Q(2k+s, 1) - Q(2k+s, E_1)
\bigr] h^2
\label{eq:fk-def}
\end{align}
where
\begin{align}
Q(\alpha,z)
&:=
\frac{2}{\pi^4}
\Realpart \biggl[
\alpha^2
\left(
\polylog_6(z) - \frac{1}{64}\polylog_6(z^2)
\right)
\notag
\\
&
\hspace{9ex}
+
\polylog_4(z) - \frac{1}{16}\polylog_4(z^2)
\biggr] ,
\end{align}
$\polylog$ is the polylogarithm \cite{nist-dig-library}
and
\begin{align}
E_1 :=
\exp \! \left(\frac{i\pi\eta_1}{\ln(b/a)}\right)
= \exp \!
\left(\frac{i\pi h \tau_1}{2\cavlength\atanh(h/2)}\right) .
\end{align}

\begin{figure}[t]
\centering
\includegraphics[width=0.45\textwidth]{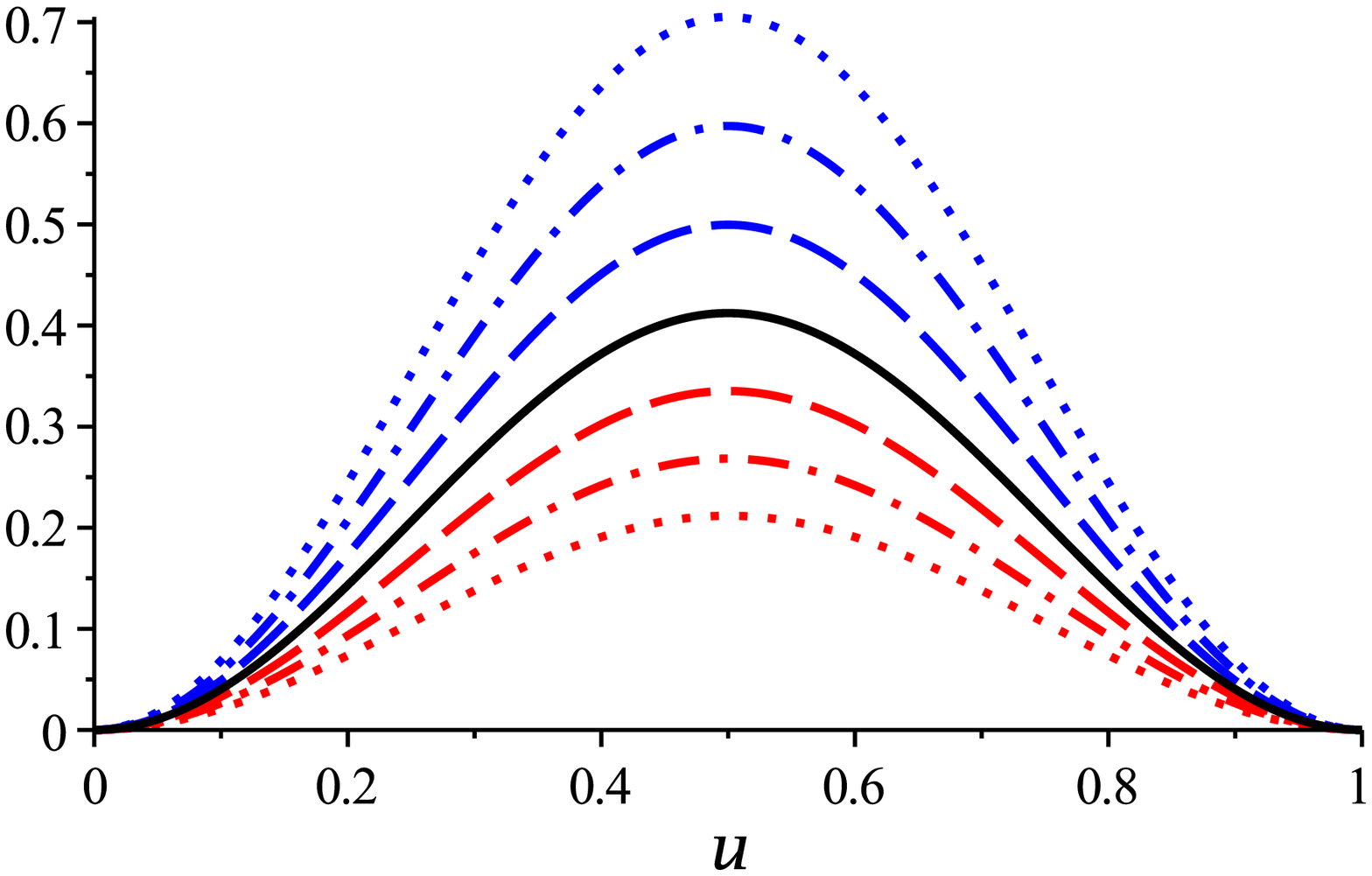}
\caption{
The plot shows $f_{k}/h^{2}$
as a function of
$u:=\tfrac12\eta_{1}/\ln(b/a) =
h \tau_1 /\bigl[4\cavlength\atanh(h/2)\bigr]$,
over the full period $0\le u \le 1$.
The solid curve (black) is for $s=0$ with $k=\pm1$.
The dashed, dash-dotted and dotted
curves are respectively for
$s=\frac14$, $s=\frac12$ and $s=\frac34$,
for $k=1$ (blue) above the solid curve and
for $k=-1$ (red) below the solid curve.
\label{fig:correction term I to III}}
\end{figure}

We see from (\ref{eq:negativity of rhoARplusminus}) that acceleration
does degrade the initially maximal entanglement,
and the degradation is determined
by the function $f_{k}$~(\ref{eq:fk-def}).
$f_{k}$~is periodic in $\tau_1$ with period
$2\cavlength {(h/2)}^{-1}\atanh(h/2)$,
which is the proper time measured at the centre of Rob's
cavity between sending and recapturing a light ray that
is allowed to bounce off each wall once.
$f_{k}$~is non-negative, and it
vanishes only at integer multiples of the period.
$f_{k}$~is not even in $k$ for generic values of~$s$, but
it is even in $k$ in the limiting case
$s=0$ in which the spectrum is symmetric
between positive and negative charges.
$f_{k}$ diverges at large $|k|$ proportionally to~$k^2$,
and the domain of validity of our
perturbative analysis is $|k| h \ll 1$.
Plots for $k=\pm1$ are shown in
Fig.~\ref{fig:correction term I to III}.

Another, potentially useful measure of entanglement for the
states at hand would be the concurrence~\cite{wootters98}.  While a
perturbative computation of the concurrence would as such be feasible,
we have verified that obtaining the leading order correction would
require expanding the states
(\ref{eq:region I single particles to order h squared})
to order~$h^4$. We have not pursued this expansion.

We now turn to nonlocality, as quantified by the
violation of the CHSH inequality \cite{chsh69,horodecki-rpm95}
\begin{align}
|\left\langle\,\mathcal{B}_{\text{CHSH}}\,\right\rangle_{\rho}|\,\leq\,2\, ,
\label{eq:chsh inequality}
\end{align}
where $\mathcal{B}_{\text{CHSH}}$ is the bipartite observable
\begin{align}
\mathcal{B}_{\text{CHSH}} :=
\mathbf{a}\cdot\mathbf{\sigma}\otimes(\mathbf{b}+\mathbf{b^{\prime}})\cdot\mathbf{\sigma}+
\mathbf{a^{\prime}}\cdot\mathbf{\sigma}\otimes(\mathbf{b}-\mathbf{b^{\prime}})\cdot\mathbf{\sigma}\,,
\label{eq:bell chsh parameter}
\end{align}	
$\mathbf{a}$, $\mathbf{a^{\prime}}$, $\mathbf{b}$ and
$\mathbf{b^{\prime}}$ are unit vectors in~$\mathbb{R}^{3}$, and
$\mathbf{\sigma}$ is the vector of the Pauli matrices. The
inequality (\ref{eq:chsh inequality}) is satisfied by all local
realistic theories, but quantum mechanics allows the left-hand side to take
values up to~$2\sqrt{2}$. The violation of (\ref{eq:chsh inequality})
is hence a sufficient
(although not necessary~\cite{friiskoehlermartinmartinezbertlmann11,smithmann11})
condition for the quantum state to be entangled.


To look for violations of (\ref{eq:chsh inequality}),
we proceed as in \cite{friiskoehlermartinmartinezbertlmann11},
noting that the maximum value of the left-hand-side in the state
$\rho$ is given by \cite{horodecki-rpm95}
\begin{align}
\left\langle\,\mathcal{B}_{\text{max}}\,\right\rangle_{\rho}\,=\,2\,\sqrt{\mu_{1}+\mu_{2}}\,,
\label{eq:chsh criterion}
\end{align}
where
$\mu_{1}$ and $\mu_{2}$ are the two largest eigenvalues of the matrix
$U(\rho)=T^{T}_{\rho}T_{\rho}$ and the elements of the correlation matrix $T=(t_{ij})$ are given by
$t_{ij}=\tr[\rho\,\sigma_{i}\otimes\sigma_{j}]$.
In our scenario
\begin{align}
U(\rho^{\pm}_{AR\pm})=
\begin{pmatrix}
1-f_{k} &   0        &  0                   \\
0       &   1-f_{k}  &  0                   \\
0       &   0        &  \tfrac{1}{4}-f_{k}  \\
\end{pmatrix}\ +\ {O}(h^{4})\,,
\label{eq:chsh matrix}
\end{align}
and working to order $h^2$ we hence find
\begin{align}	
\left\langle\,\mathcal{B}_{\text{max}}\,\right\rangle_{\rho^{\pm}_{AR\pm}}\,=\,
2\,\sqrt{2} \left(\,1 - \tfrac{1}{2} f_k \right)\,.
\label{eq:chsh violation for rhoARplusminus}
\end{align}
The acceleration thus degrades the initially maximal
violation of the CHSH inequality,
and the degradation is again determined by the function~$f_k$.

\subsection{\label{subsec:particle antiparticle entanglement}Entanglement
between opposite charges}

We finally turn to the entanglement between opposite charges in the state
$\left|\,\chi^{\pm}_{\text{init}}\,\right\rangle_{AR}$~(\ref{eq:initial particle antiparticle entangled state}).

Expressing the density matrix in the Region $\mathrm{I\!I\!I}$ basis,
tracing over Rob's unobserved modes and working perturbatively to
order~$h^2$, we find that the only nonvanishing elements of the
reduced density matrix are within a $6\times6$ block.  Partially
transposing Rob's subsystem replaces the last lines in
(\ref{eq:par-tr k,kprime 1- 1-}) and (\ref{eq:par-tr k,kprime 1+ 1+})
by their respective conjugates and shifts the particle-antiparticle
off-diagonals (\ref{eq:par-tr k,kprime 1+ 1-}) away from the
diagonal. The only nonvanishing elements of the partial transpose are
thus within an $8\times8$ block, which decomposes further into two
$3\times3$ blocks that correspond respectively to
(\ref{eq:par-tr k,kprime 1- 1-}) and (\ref{eq:par-tr k,kprime 1+ 1+}) and the
$2\times2$ block
\begin{align}
\pm\tfrac{1}{2}
\begin{pmatrix}
0
&
\hspace*{-6ex}
G_{k}G_{k^{\prime}}\bigl|\mathcal{A}_{kk^{\prime}}^{(1)}\bigr|^{2} +
\mathcal{A}_{kk}\mathcal{A}_{k^{\prime}k^{\prime}}
\\[1.5mm]
G_{k}^{\,*}G_{k^{\prime}}^{\,*}\bigl|\mathcal{A}_{kk^{\prime}}^{(1)}\bigr|^{2} +
\mathcal{A}_{kk}^{*}\mathcal{A}_{k^{\prime}k^{\prime}}^{*}
&
\hspace*{-6ex}0\\
\end{pmatrix} ,
\label{eq:nonzero two by two block}
\end{align}
where the off-diagonal components are kept only to
order $h^2$ in their small $h$ expansion~(\ref{eq:Akk times Akprimekrprime expansion}).

The only negative eigenvalue comes from the
$2\times2$ block~(\ref{eq:nonzero two by two block}).
We find that the negativity is given by
\begin{align}
\mathcal{N}[\rho_{\chi}^{\pm}]
&=
\tfrac{1}{2} \,-\,
\tfrac{1}{4}\,\sum\limits_{p\neq k^{\prime}}\bigl|\mathcal{A}_{kp}^{(1)}\bigr|^{2}\,-\,
\tfrac{1}{4}\,\sum\limits_{p\neq k}\bigl|\mathcal{A}_{k^{\prime}p}^{(1)}\bigr|^{2}
\notag
\\
&=
\tfrac12
\,-\,
\tfrac14
\left(
f_k + f_{k'}
\right)
+ \tfrac12
\bigl|E_1^{k-k'} - 1\bigr|^2
\bigl| A_{k k'}^{(1)}\bigr|^{2}  .
\label{eq:negativity of rhochiplusminus}
\end{align}
The entanglement is hence again degraded by the acceleration, and the
degradation has the same periodicity in $\tau_1$ as in the cases
considered above. The degradation now depends however on $k$ and $k'$ not just
through the individual functions $f_k$ and $f_{k'}$ but also through
the term proportional to $\bigl| A_{k k'}^{(1)}\bigr|^{2}$
in~(\ref{eq:negativity of rhochiplusminus}): this interference term is
nonvanishing iff $k$ and $k'$ have different parity, and when it is
nonvanishing, it diminishes the degradation effect. In the charge-symmetric
special case of $s=0$ and $k=-k'$, the degradation coincides with that
found in (\ref{eq:negativity of rhoARplusminus}) for the two-mode
states~(\ref{eq:initial pure state rob both}).

\section{\label{sec:onewaytrip}One-way journey}

Our analysis for the Rob trajectory that comprises Regions
$\mathrm{I}$, $\mathrm{I\!I}$ and~$\mathrm{I\!I\!I}$ can be
generalised in a straightforward way to any trajectory obtained by
grafting inertial and uniformly-accelerated segments, with arbitrary
durations and proper accelerations. The only delicate point is that the
phase conventions of our mode functions distinguish the left boundary
of the cavity from the right boundary, and in Sec.~\ref{subsec:I-to-II}
we set up the Bogoliubov transformation from Minkowski to Rindler
assuming that the acceleration is to the right. It follows
that the Bogoliubov transformation from Minkowski to
leftward-accelerating Rindler is obtained from that in
Sec.~\ref{subsec:I-to-II} by inserting the appropriate phase
factors, $A_{mn} \to {(-1)}^{m+n} A_{mn}$, and in the expansions
(\ref{eq:bogo coeff pert}) this amounts to the replacement $h \to -h$.

As an example, consider the Rob cavity trajectory that starts
inertial, accelerates to the right for proper time $\tau_1$ as above,
coasts inertially for proper time $\tau_2$ and finally performs a
braking manoeuvre that is the reverse of the initial acceleration,
ending in an inertial state that has vanishing velocity with respect
the initial inertial state. Denoting the mode functions in the final
inertial state by ${\widetilde{\!\widetilde\psi}}_n$, and writing
\begin{align}
\label{eq:onewaychange-of-basis}
{\widetilde{\!\widetilde\psi}}_m
= \sum\limits_{n}\mathcal{B}_{mn}\,\psi_{n}
\ ,
\end{align}
we find
\begin{align}
\bigl| \mathcal{B}_{mn}^{(1)}\bigr|^{2}
=
\bigl|E_1^{m-n} - 1\bigr|^2
\bigl|{(E_1 E_2)}^{m-n} - 1\bigr|^2
\bigl| A_{mn}^{(1)}\bigr|^{2}
\end{align}
where
$E_2 := \exp(i\pi \tau_2/\cavlength)$.
For the two-mode initial
states
$\left|\,\phi^{\pm}_{\text{init}}\,\right\rangle_{AR+}$ and
$\left|\,\phi^{\pm}_{\text{init}}\,\right\rangle_{AR-}
$~(\ref{eq:initial pure state rob both}),
the negativity and the maximum violation of the CHSH inequality
hence read respectively
\begin{subequations}
\begin{align}
\mathcal{N}[\rho^{\pm}_{AR\pm}]
& \,=\,
\tfrac{1}{2} \Bigl( 1 - {\widetilde{\!\widetilde f}}_k \Bigr) ,
\label{eq:onewaynegativity of rhoARplusminus}
\\
\left\langle\,\mathcal{B}_{\text{max}}\,\right\rangle_{\rho^{\pm}_{AR\pm}}
& \,=\,
2\,\sqrt{2} \Bigl(\,1 - \tfrac{1}{2} {\widetilde{\!\widetilde f}}_k\Bigr)\,,
\label{eq:onewaychsh violation for rhoARplusminus}
\end{align}
\end{subequations}
where
\begin{align}
{\widetilde{\!\widetilde f}}_k
&
=
\sum\limits_{p=-\infty}^{\infty}
\bigl| \mathcal{B}_{kp}^{(1)}\bigr|^{2}
\notag
\\[1ex]
& =
2 \bigl[
2Q(2k+s,1) - 2Q(2k+s,E_1) + Q(2k+s,E_2)
\notag
\\
&
\hspace{5ex}
- 2Q(2k+s,E_1E_2) +
Q(2k+s,E_1^2 E_2)
\bigr] h^2 \ .
\label{eq:onewayfk-def}
\end{align}
The negativity in the state
$\left|\,\chi^{\pm}_{\text{init}}\,\right\rangle_{AR}$
(\ref{eq:initial particle antiparticle entangled state}) reads
\begin{align}
\mathcal{N}[\rho_{\chi}^{\pm}]
&=
\tfrac12
\,-\,
\tfrac14
\Bigl(
{\widetilde{\!\widetilde f}}_k
+ {\widetilde{\!\widetilde f}}_{k'}
\Bigr)
\notag
\\
&
\hspace{2ex}
+ \tfrac12
\bigl|E_1^{k-k'} - 1\bigr|^2
\bigl|{(E_1E_2)}^{k-k'} - 1\bigr|^2
\bigl| A_{k k'}^{(1)}\bigr|^{2}  .
\label{eq:onewaynegativity of rhochiplusminus}
\end{align}

\begin{figure}[t]
\centering
\vspace*{-4ex}%
\includegraphics[width=0.45\textwidth]{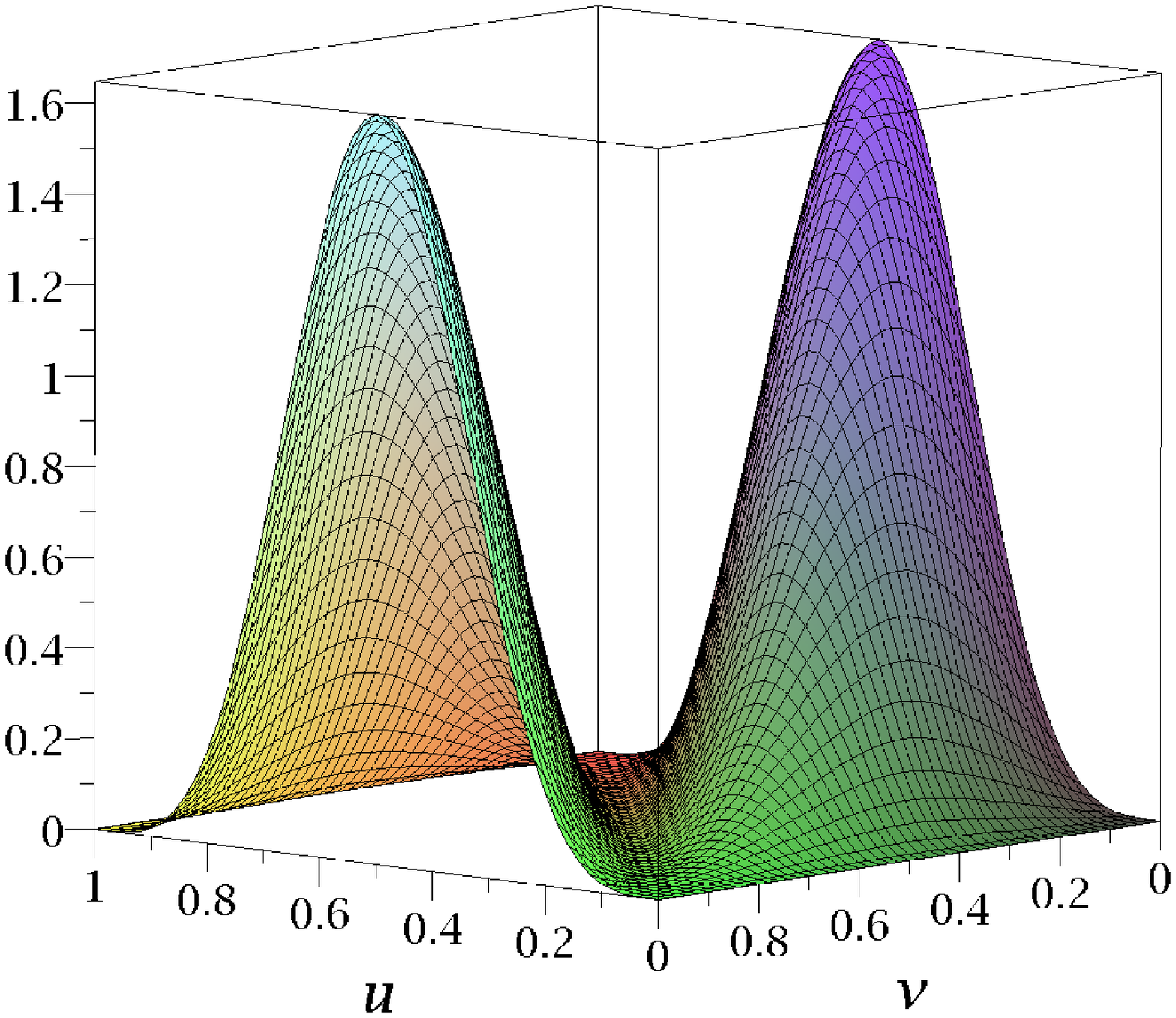}
\caption{
The plot shows ${\widetilde{\!\widetilde f}}_k$
as a function of
$u:= h \tau_1 /\bigl[4\cavlength\atanh(h/2)\bigr]$
and
$v:= \tau_2 /(2\cavlength)$
over the full period $0\le u \le 1$ and  $0\le v \le 1$,
for $s=0$ and $k=1$.  Note the zeroes at $u\equiv0 \mod 1$ and at
$u+v \equiv 0 \mod 1$.
\label{fig:onewaytripfigure}}
\end{figure}

The degradation caused by acceleration is thus again periodic in
$\tau_1$ with period $2\cavlength {(h/2)}^{-1}\atanh(h/2)$, and it is
periodic in $\tau_2$ with period~$2\cavlength$.  The degradation
vanishes iff $E_1=1$ or $E_1 E_2=1$, so that any degradation caused by
the accelerated segments can be cancelled by fine-tuning the duration
of the inertial segment, to the order $h^2$ in which we are working.
A~plot of ${\widetilde{\!\widetilde f}}_k$
is shown in Fig.~\ref{fig:onewaytripfigure}.

\section{\label{sec:conclusion}Conclusions}

We have analysed the entanglement degradation for a massless Dirac
field between two cavities in $(1+1)$-dimensional Minkowski spacetime, one
cavity inertial and the other moving along a trajectory that
consists of inertial and uniformly accelerated segments. Working in
the approximation of small accelerations but arbitrarily long travel
times, we found that
the degradation is qualitatively similar to that found in
\cite{bruschifuenteslouko11} for a massless scalar field with
Dirichlet boundary conditions. The degradation is
periodic in the durations of the individual inertial and accelerated
segments, and we identified a travel scenario where the degradation
caused by accelerated segments can be undone by fine-tuning the
duration of an inertial segment. The presence of charge allows however
a wider range of initial states of interest to be analysed. As an
example, we identified a state where the entanglement degradation
contains a contribution due to interference between excitations of
opposite charge.

Compared with bosons, working in a fermionic Fock space led both to
technical simplifications and to technical complications. A~technical
simplification was that the relevant reduced density matrices act in a
lower-dimensional Hilbert space because of the fermionic statistics,
and this made it possible to quantify the entanglement not just in
terms of the negativity but also in terms of the CHSH inequality. It
would further be possible to investigate the concurrence, although
doing so would require pushing the perturbative low-acceleration
expansion to a higher order than we have done in this paper.

A~technical complication was that when the boundary conditions at the
cavity walls were chosen in an arguably natural way that preserves
charge conjugation symmetry, the spectrum contained a zero mode. This
zero mode could not be consistently omitted by hand, but we were able
to regularise the zero mode by treating the charge-symmetric boundary
conditions as a limiting case of charge-nonsymmetric boundary
conditions. All our entanglement measures remained manifestly well
defined when the regulator was removed.

Another technical complication occurring for fermions is the
ambiguity \cite{monteromartinmartinez11} in the choice of the basis of
the two-fermion Hilbert space
in (\ref{eq:par-tr k,kprime all}). An alternative valid choice of basis is obtained by reversing the order of
the single particle kets in (\ref{eq:par-tr k,kprime all}), which amounts
to a change of the signs in the off-diagonal elements of (\ref{eq:par-tr k,kprime 1- 1-}) and (\ref{eq:par-tr k,kprime  1+ 1+}). While our treatment does not remove this ambiguity, all of our results for the entanglement and the nonlocality of these states are independent of the chosen convention.

Our analysis contained two significant limitations. First, while our
Bogoliubov transformation technique can be applied to
arbitrarily complicated graftings of inertial and
uniformly accelerated cavity trajectory segments,
the treatment is perturbative in
the accelerations and hence valid only in the small acceleration
limit. We were thus not able to address the large acceleration limit,
in which striking qualitative differences between bosonic and
fermionic entanglement have been found for field modes that are not
confined in
cavities~\cite{fuentesschullermann05,alsingfuentesschullermanntessier06,%
  bruschiloukomartinmartinezdraganfuentes10,%
  martinmartinezfuentes11,friiskoehlermartinmartinezbertlmann11}.

Second, a massless fermion in a $(1+1)$-dimensional cavity is unlikely
to be a good model for systems realisable in a laboratory. A~fermion
in a linearly-accelerated rectangular cavity in $(3+1)$ dimensions can
be reduced to the $(1+1)$-dimensional case by separation of variables,
but for generic field modes the transverse quantum numbers then
contribute to the effective $(1+1)$-dimensional mass; further, any
foreseeable experiment would presumably need to use fermions that have
a positive mass already in $(3+1)$ dimensions before the reduction. It
would be possible to analyse our $(1+1)$-dimensional system for a
massive fermion, and we anticipate that the mass would enhance the
magnitude of the entanglement degradation as in the bosonic
situation~\cite{bruschifuenteslouko11}. A~detailed analysis of a
massive fermion could become of experimental interest if guided by
insights as to how a massive fermion might be confined to a cavity in
a concrete laboratory setting.

We started this paper by emphasising that a cavity localises the
quantum degrees of freedom in the worldtube of the cavity, and our
assumption of inertial initial and final trajectory segments localises
the acceleration effects in a finite interval of the cavity's proper
time. We should perhaps end by emphasising that we are not attempting
to localise measurements of the field at more precise spacetime
locations within the worldtube of the cavity, and we are hence not
proposing cavities as a fundamental solution to the open conceptual
issues of a quantum measurement theory in relativistic
spacetime~\cite{sorkin-impossible}. A~cavity can however reduce the
measurement ambiguities from, say, megaparsecs to centimetres, which
may well suffice to resolve the conceptual issues in specific
experimental settings of interest, gedanken or otherwise.

\begin{acknowledgments}
We thank Gerardo Adesso,
Reinhold Bertlmann,
Fay Dowker,
Ivette Fuentes,
Beatrix C. Hiesmayr,
Marcus Huber
and
Eduardo Mart\'{i}n-Mart\'{i}nez
for helpful discussions and comments.
N.~F. acknowledges support from EPSRC
[CAF Grant No.~EP/G00496X/2 to I.~Fuentes]
and the $\chi$-QEN collaboration.
J.~L. was supported in part by STFC.
\end{acknowledgments}

\end{document}